\documentclass[pra,reprint]{revtex4-1}

\usepackage{amssymb}
\usepackage{amsmath}
\usepackage{graphicx}

\begin{document}

\title{
  Light-matter interaction at rough surfaces: \\
  a morphological perspective on laser-induced periodic surface structures
}

\author{Vladimir Yu. Fedorov}
\email{v.y.fedorov@gmail.com}
\affiliation{Laboratoire Hubert Curien, Université Jean Monnet, Saint-Etienne, France}

\author{Jean-Philippe Colombier}
\email{jean.philippe.colombier@univ-st-etienne.fr}
\affiliation{Laboratoire Hubert Curien, Université Jean Monnet, Saint-Etienne, France}

\date{\today}

\begin{abstract}
We use ab-initio electromagnetic simulations to investigate light absorption by rough
surfaces in the context of the formation of laser-induced periodic surface structures.
Our approach involves modeling a realistic rough surface using a statistical description of
its continuous height distribution via a corresponding correlation function.
We study the influence of incident light polarization and various statistical properties of
surface roughness, such as root-mean-square height and correlation length, on the
distribution of absorbed laser energy.
By analyzing light absorption in different layers of the surface selvedge, we elucidate how
different features of surface morphology influence the shape of the resulting periodic
surface structures.
We show that circularly polarized laser pulses are highly sensitive to initial or
progressively developing asymmetries in surface roughness.
\end{abstract}

\maketitle

\section{Introduction}
When an intense ultrashort laser pulse hits the surface of a solid, it triggers a chain of
events, each unfolding on a specific time scale.
In the initial stage occurring on a femtosecond time scale, within the pulse duration, a
strong laser field excites electrons in the material; light scatters by surface roughness
and, in case of metallic media, generates surface plasmons.
Then, on a picosecond time scale, after the passage of the laser pulse, the electron-phonon
thermalization stage begins.
At this stage, hot excited electrons, colliding with atoms of the material, transfer their
energy to the lattice, thereby heating the material.
Next, on a time scale from a few picoseconds to nanoseconds, hydrodynamic processes come
into play.
During this period, material melting, cavitation and explosive boiling occur, accompanied by
convection and hydrodynamic instabilities that ultimately shape the surface as the material
solidifies.
Finally, approximately a millisecond later (corresponding to a typical kilohertz laser
repetition rate) the succeeding pulse within the pulse train arrives and triggers the same
chain of events again.
During this iterative process, each subsequent laser pulse interacts with the surface that
has been modified by the previous pulses.
This interaction forms a feedback loop, which, after multiple cycles, leads to the emergence
of periodic ripples known as laser-induced periodic surface structures
(LIPSS)~\cite{Bonse2017,Graef2020,Bonse2020,Rudenko2023}.

The formation of LIPSS is a universal phenomenon: they can be produced on almost all types of
materials (dielectrics, semiconductors, metals).
By fine-tuning parameters of LIPSS, we can change surface properties and thereby
functionalize the irradiated material.
For example, since LIPSS act as a diffraction grating, by adjusting their period we can
select which wavelengths of light will be reflected and thus change the color of
materials~\cite{Vorobyev2008,Dusser2010,Li2015}.
In addition, with LIPSS we can alter the wettability of the surface and create hydrophobic
or hydrophilic coatings~\cite{Zorba2008,Kietzig2009,Kunz2018}.
Also, using LIPSS we can control the friction coefficient of materials~\cite{Yu1999,
Bonse2018,Kunz2020} or impart antibacterial properties to their surfaces~\cite{Jalil2020,
Sotelo2023,Outon2024}.

Depending on the period, LIPSS can be divided in two main classes: LIPSS whose period is
close to the laser wavelength are referred to as low spatial frequency LIPSS (LSFLs) and
those whose period is significantly less than the laser wavelength are referred to as high
spatial frequency LIPSS (HSFLs)~\cite{Bonse2017,Graef2020,Rudenko2023}.
Among the two, HSFLs are of particular interest because they enable surface nanostructuring
with a resolution well below the optical diffraction limit~\cite{Stoian2020}.
Interestingly, to date HSFLs have only been achieved using ultrashort femtosecond or
picosecond laser pulses.

The orientation of LIPSS depends on material properties and can be either parallel or
perpendicular to the laser polarization.
The determination of which orientation prevails occurs at the initial stage of laser-matter
interaction where anisotropy of light scattering by inevitable surface defects leads to
symmetry breaking.
The distribution of scattered light intensity, and consequently, the spatial modulation of
local energy (fluence) imprinted on the material surface through absorption, is dictated by
the surface roughness.
The inhomogeneous distribution of absorbed light energy acts as a seed, setting on the
amplification process through the inter-pulse feedback loop, thereby shaping the final
morphology of LIPSS.
Thus, an accurate description of surface morphology and its influence on scattering and
absorption of incoming light is of paramount importance for understanding the physics
underlying the LIPSS generation.

Although there are a number of experiments on the influence of surface quality on LIPSS
formation, a thorough and comprehensive study in this area is clearly lacking.
It was shown that the surface roughness directly affects the period of LIPSS and thus can
alter certain surface functionality~\cite{Fuentes2019,Sotelo2023}.
The formation of consistent LIPSS can be significantly enhanced by polishing the surface in
the direction perpendicular to the expected orientation of LIPSS~\cite{Ardron2014}.
In contrast, deposition of thin films through pulse vapor deposition provides a flexible
method for preventing aligned scratch and adjusting the initial nanostructures of the
topography, which in turn influences the feedback process significantly~\cite{Prudent2021}.

In theoretical studies, the importance of surface roughness in explaining the origin of
LIPSS has become clear starting from the pioneering works of Sipe et al.~\cite{Sipe1983} who
linked the appearance of LIPSS with the interference of the incident laser radiation with
surface electromagnetic waves generated by scattering at the rough surface.
Later, the crucial role of light interaction with surface inhomogeneities in the formation
of LIPSS has been confirmed using ab-initio electromagnetic simulations, free from the
limitations of the Sipe's theory~\cite{Skolski2012,Skolski2014,Zhang2015,Zhang2020}.
However, in all of these simulations the surface roughness was considered simply as a
discrete binary mask~--- a collection of identical randomly distributed sub-wavelength
scatterers on top of a flat surface.
Meanwhile, any realistic rough surface possesses a continuous distribution of heights, which
requires description in terms of statistical properties like the height probability
distribution and the height correlation function.
In this paper we apply such statistical description of surface roughness in order to
simulate the scattering and subsequent absorption of ultrashort laser pulses by rough
surfaces.
We study how various statistical properties of the surface roughness affect the distribution
of the absorbed light energy.
In particular, we demonstrate that different layers of surfaces with a continuous height
distribution scatter light in different directions.
This observation allows us to trace which surface layers are responsible for the formation
of certain features of the LIPSS shape.
Finally, we explore how light polarization affects the distribution of absorbed energy on
rough surfaces.
We show that circularly polarized laser pulses allow to detect initial or gradually
developing asymmetries in surface roughness.

\section{Model of the surface roughness}
A rough surface can be described by the roughness function $R(x,y)$, which defines the
random deviations of the surface height relative to some reference plane~\cite{Ogilvy1992}.
With its help, the statistical properties of the surface roughness can be expressed using
the correlation function $C(X,Y)$ of the form
\begin{align}
  C(X,Y) = \frac{\langle R(x,y)R(x+X,y+Y) \rangle}{\sigma^2},
\end{align}
where $\sigma=\sqrt{\langle R^2 \rangle}$ is the root-mean-square (rms) height and
$\langle\dots\rangle$ denotes the spatial averaging.
The correlation function $C(X,Y)$ describes the spatial coherence between different surface
points separated by a distance $d=\sqrt{X^2+Y^2}$.
Typical cases are the Gaussian correlation function
\begin{align} \label{eq:cgauss}
  C(X,Y) = \sigma^2 \exp\left(-\frac{X^2 + Y^2}{\xi^2}\right)
\end{align}
and the exponential correlation function
\begin{align} \label{eq:cexp}
  C(X,Y) = \sigma^2 \exp\left(-\frac{\sqrt{X^2 + Y^2}}{\xi}\right)
\end{align}
with $\xi$ being the correlation length.

In turn, the power spectral density $W(K_x,K_y)$ characterizes the average distribution of
the spatial frequency components of randomly fluctuated surface profile.
It is related to the correlation function $C(X,Y)$ by the Fourier transform as follows:
\begin{align}
  & W(K_x,K_y) = \notag \\
  & \qquad \frac{1}{(2\pi)^2}
    \iint_{-\infty}^{\infty} C(X,Y) e^{-i(K_xX + K_yY)} dX dY,
\end{align}
where $K_x$ and $K_y$ are the spatial frequencies.

To generate a rough surface numerically, we introduce a spectral function $F(K_x,K_y)$ which
we obtain from the power spectral density $W(K_x,K_y)$ by modifying its amplitude and phase
by a sequence of normally distributed random numbers~\cite{Thorsos1988,Mack2013}.
This new function allows us to calculate the roughness $R(x,y)$ using the following inverse
Fourier transform:
\begin{align}
  R(x,y) = \iint_{-\infty}^{\infty} F(K_x,K_y) e^{i(K_xx + K_yy)} dK_x dK_y.
\end{align}

Function $F(K_x,K_y)$ is constructed as follows~\cite{Thorsos1988,Mack2013}.
For all positive frequencies $K_x>0$, $K_y>0$ we define $F(K_x,K_y)$ as
\begin{align}
  F(K_x,K_y) = \sqrt{L_xL_y W(K_x,K_y)} ~ \frac{\eta_1 + i\eta_2}{\sqrt{2}}
\end{align}
and for zero and Nyquist frequencies ($K_x=0$, $K_y=0$ or $K_x=\max(K_x)$, $K_y=\max(K_y)$)
as
\begin{align}
  F(K_x,K_y) = \sqrt{L_xL_y W(K_x,K_y)} ~ \eta_3.
\end{align}
Here $L_x$ and $L_y$ are the lengths of $x$ and $y$ numerical domains while each of
$\eta_1$, $\eta_2$ and $\eta_3$ denotes a sequence of normally distributed random numbers in
the range $[0,1]$ with zero mean and unity standard deviation.
Since the surface roughness $R(x,y)$ must be a real valued function, $F(K_x,K_y)$ have to
satisfy the following additional conditions:
\begin{align}
  F(-K_x,-K_y) & = F^*(K_x,K_y), \\
  F(-K_x,K_y) & = F^*(K_x,-K_y).
\end{align}
In other words, $F(K_x,K_y)$ must be "conjugate symmetric" about the origin, i.e. the
reflection of $F(K_x,K_y)$ at any point relative the origin must be its complex conjugate.

Thus, in order to obtain the surface roughness function $R(x,y)$ with the desired
statistical properties we first select a correlation function $C(X,Y)$.
Then, using the Fourier transform, we calculate the power spectral density $W(K_x,K_y)$.
After, we construct the function $F(K_x,K_y)$ and, finally, applying the inverse Fourier
transform to $F(K_x,K_y)$, we calculate $R(x,y)$.

\begin{figure}
  \includegraphics[width=\columnwidth]{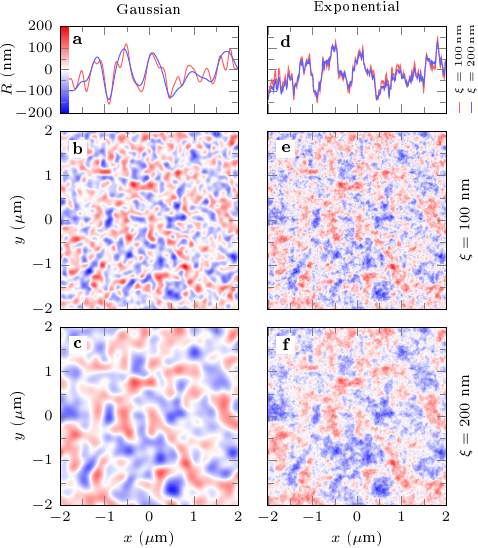}
  \caption{
    Surface roughness $R(x,y)$ for the Gaussian (a,b,c) and exponential (d,e,f) correlation
    functions with the rms height $\sigma=50$~nm and two different correlation lengths
    $\xi=100$ (b,e) and 200~nm (c,f).
    The line plots (a,d) show the corresponding cross-sections at $y=0$.
    \label{fig:rough}
  }
\end{figure}

Figure~\ref{fig:rough} shows a typical surface roughness $R(x,y)$ generated by the above
approach for the Gaussian, Eq.~\eqref{eq:cgauss}, and exponential, Eq.~\eqref{eq:cexp},
correlation functions.
For the both cases we use the same seed for the random numbers, the rms height
$\sigma=50$~nm and consider two different correlation lengths, $\xi=100$ and 200~nm.
In Figs.~\ref{fig:rough}(a,b,c) we see that for the Gaussian correlation function the
correlation length $\xi$ determines the average scale of surface inhomogeneities.
In turn, Figs.~\ref{fig:rough}(d,e,f) show that for the exponential correlation function
a faster decay of the spatial coherence results in a surface with very rapidly varying
small-scale perturbations superimposed on the Gaussian roughness.
In this case the correlation length $\xi$ determines the amount of these perturbations and
the sharpness of their peaks.

\section{Numerical model}
To study the interaction of ultrashort laser pulses with rough surfaces we numerically solve
the following system of Maxwell equations using the finite-difference time-domain method
(FDTD)~\cite{Taflove2005}:
\begin{align}
  \vec{\nabla}\times\vec{E} = -\frac{\partial\vec{B}}{\partial t}, \quad
  \vec{\nabla}\times\vec{H} = \frac{\partial\vec{D}}{\partial t},
\end{align}
where $\vec{E}(x,y,z,t)$ and $\vec{H}(x,y,z,t)$ are the electric and magnetic field vectors,
$\vec{B}=\mu_0\vec{H}$, with $\mu_0$ being the vacuum permeability, and
$\widetilde{D}=\varepsilon_0\varepsilon(\omega)\widetilde{E}$, where
$\widetilde{\hphantom{E}}$ means the temporal spectrum, while $\varepsilon_0$ is the vacuum
permittivity and $\varepsilon(\omega)$ is the frequency-dependent permittivity of the
medium.

In our simulations we use the calculation grid with the sizes $L_x=L_y=12.2$~$\mu$m and
$L_z=0.9$~$\mu$m in the $x$, $y$ and $z$ directions, respectively, with the corresponding
step sizes $\Delta x=\Delta y=10$~nm and $\Delta z=5$~nm.
To avoid nonphysical reflections, at each end of the grid we place convolutional perfectly
matched layers of 0.1~$\mu$m thickness.

As the incoming laser radiation we use a plane wave propagating along $z$ direction, whose
temporal envelope $A(t)$ is given by a one period of $\sin^2$ function:
$A(t)=\sin^2(\pi t/4\tau_0)$, where $\tau_0$ is the laser pulse duration.
In the case of a laser pulse polarized in the $x$ direction, we excite the $x$ component of
the electric field: $E_x(t)=A(t)\cos(\omega_0t)$, where $\omega_0=2\pi c_0/\lambda_0$ is the
central frequency with $\lambda_0$ and $c_0$ being the central wavelength and the speed of
light in vacuum.
To model a circularly polarized laser pulse we also add the $y$ component of the field with
the relative phase shift of $\pi/2$: $E_y(t)=A(t)\sin(\omega_0t)$.
To preserve the pulse energy, for circularly polarized pulses we divide $E_x$ and $E_y$ by
$\sqrt{2}$.
In our simulations we use ultrashort laser pulses with the duration $\tau_0=50$~fs and the
central wavelength $\lambda_0=1.03$~$\mu$m.

At the distance of 0.6~$\mu$m below the radiation plane (as measured relative to the
reference plane of the roughness function $R(x,y)$), we place a semi-infinite stainless
steel medium with a rough surface.
To model the dispersive response of stainless steel we apply the auxiliary differential
equation method~\cite{Taflove2005} and assume the Drude permittivity
$\varepsilon(\omega) = 1 - \omega_p^2/(\omega^2 + i\omega\gamma)$ with the plasma frequency
$\omega_p=19.2\times10^{15}$~1/s and the damping rate
$\gamma=9.15\times10^{15}$~1/s~\cite{Rudenko2019}.
Under these parameters the complex refractive index $n=n'+in''$ of stainless steel at
central frequency $\omega_0$ has the real and imaginary parts equal to $n'=3.02$ and
$n''=3.51$.
The corresponding skin depth is $l=1/(2n''\omega_0/c_0)=23.34$~nm.

In our studies we are mainly interested in the distribution of the laser pulse energy
absorbed on the surface: the surface areas that have absorbed sufficiently large amount of
energy will later be extruded after the thermalization and hydrodynamic stages of the
laser-matter interaction.
Therefore any asymmetry in the distribution of the absorbed energy serves as a seed for
LIPSS growth.
To estimate the amount of energy absorbed on the surface let us first consider the power $P$
delivered by the laser pulse per unit time per unit volume of the material:
$P=\vec{E}\cdot\vec{J}$, where $\vec{E}$ is the electric field of the laser pulse and
$\vec{J}$ is the current of the excited electrons~\cite{Griffiths1999}.
For the current $\vec{J}$ we can write the Omh's law in the form $\vec{J}=\sigma\vec{E}$,
where the conductivity $\sigma$, taken at the central frequency $\omega_0$, can be expressed
through the imaginary part of the permittivity $\varepsilon=\varepsilon'+i\varepsilon''$ as
$\sigma=\varepsilon_0\omega_0\varepsilon''$.
Finally, we can write $\varepsilon''=2n'n''$, where $n'$ and $n''$ are the real and
imaginary parts of the refractive index $n=n'+in''=\sqrt{\varepsilon}$.
Combining all these expressions together, we obtain $P=2\varepsilon_0n'n''\omega_0E^2$.
Then, to obtain the spatial distribution of the energy $Q$ delivered by the laser pulse per
unit volume we have to integrate $P$ over time:
$Q(x,y,z)=2\varepsilon_0n'n''\omega_0\int_{-\infty}^\infty E(x,y,z,t)^2dt$.
Thus, the electric field $\vec{E}$, obtained during the FDTD simulations, allows us to
calculate the desired distribution of the absorbed energy $Q$ in the entire material below
its surface.

\section{Results and discussion}
\begin{figure}
  \includegraphics[width=\columnwidth]{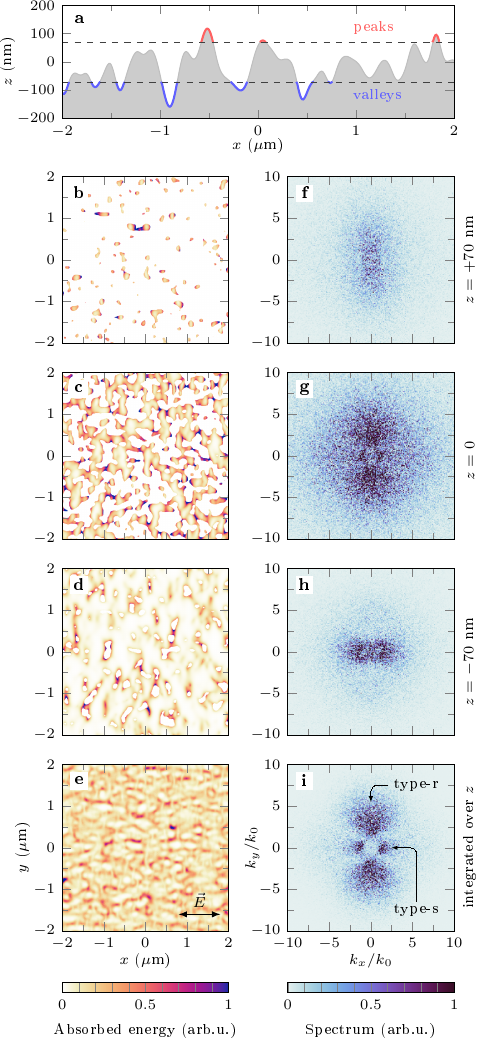}
  \caption{
    (a) The $y=0$ cross-section of the rough surface.
    The dashed lines mark $z=\pm70$~nm levels.
    The surface has the Gaussian correlation function with the rms height $\sigma=50$~nm and
    the correlation length $\xi=100$~nm.
    (b,c,d) The distribution of absorbed energy $Q$ in three surface layers $z=+70$, 0,
    -70~nm.
    (e) The distribution of absorbed energy $Q_\text{i}$ integrated over all surface layers
    along $z$; the double arrow shows the direction of laser polarization.
    (f--i) The spatial power spectra of the corresponding $Q$ distributions.
    The arrows in (i) mark the characteristic spectral patterns known as "type-r" and
    "type-s" features.
    \label{fig:layers}
  }
\end{figure}

First, let us study the distribution of the absorbed energy $Q(x,y,z)$ in different layers
of the rough surface exposed to an $x$-polarized laser pulse.
To model the rough surface we take the Gaussian correlation function with the rms height
$\sigma=50$~nm and the correlation length $\xi=100$~nm.
Figure~\ref{fig:layers}(a) shows the $y=0$ cross-section of the corresponding roughness
function $R(x,y)$.
In turn, Figs.~\ref{fig:layers}(b,c,d) show the $xy$ slices of $Q$ in three different
surface layers $z=+70$, 0, -70~nm (see the dashed lines in Fig.~\ref{fig:layers}(a)).
Note that in all absorbed energy maps presented in the paper, in order to better examine
small details, we plot only the central part of the calculation grid with the size 4 by
4~$\mu$m, while the full grid size is 12 by 12~$\mu$m.
The white areas in Figs.~\ref{fig:layers}(b,c,d) correspond to the regions outside of the
material, where $Q=0$.
We observe that, as a result of near field enhancement, the maxima of the absorbed energy
$Q$ are concentrated in the vicinity of the sharp edges of the surface.

Figures~\ref{fig:layers}(f,g,h) show the spatial spectra of the $Q$ slices in the
spatial-frequency coordinates $k_x$ and $k_y$ normalized by the wave number
$k_0=\omega_0/c_0$.
These spatial spectra give us an idea of typical sizes and orientation of the absorbed
energy spots.
The spectral components of $Q$ at spatial frequencies close to or less than $k_0$ correspond
to the large spots whose characteristic sizes are comparable to or exceeding the laser
wavelength $\lambda_0$.
They are responsible for the formation of LSFLs.
In contrast, the spectral components of $Q$ at spatial frequencies significantly exceeding
$k_0$ correspond to the small-scale sub-wavelength spots responsible for generation of
HSFLs.
For example, in Figs.~\ref{fig:layers}(f,g,h) we see that the highest frequency components
of the spectra lie in the region of $5k_0$.
Therefore, we can conclude that in all three considered layers the distribution of the
absorbed energy $Q$ has morphological features with a minimum size approximately five times
smaller than the laser wavelength $\lambda_0$.

Figures~\ref{fig:layers}(f,g,h) also show that the spectral distributions of the absorbed
energy $Q$ in different layers of the surface are not isotropic.
The spectrum corresponding to $z=+70$~nm layer is elongated in the $k_y$ direction
(see Fig.~\ref{fig:layers}(f)), while the spectrum corresponding to $z=-70$~nm layer is
elongated in the perpendicular $k_x$ direction (see Fig.~\ref{fig:layers}(h)); the spectrum
for $z=0$ layer represents the transition between the two previous ones.
Therefore, taking into account that smaller features in the spatial domain correspond to
higher spectral frequencies, we can conclude that the spots of absorbed energy $Q$ in
different surface layers have, in average, different orientation.
At the top edge of the surface the spots of the absorbed energy are mainly stretched in the
$x$ direction, parallel to the laser polarization, while at the bottom edge they line up in
the perpendicular $y$ direction.
In order to explain this result we refer to our recent study where we showed that a single
nanobump or nanocavity, irradiated by linearly polarized laser light, both have an
asymmetric scattering pattern similar to that of a dipole~\cite{Rudenko2019}.
In particularly, we showed that nanobumps and nanocavities scatter light predominantly in
mutually perpendicular directions: nanobumps along the laser polarization and nanocavities
in the perpendicular one (see Fig.~3 in~\cite{Rudenko2019}).
In turn, as shown in Fig.~\ref{fig:rough}(a), any rough surface can be considered as a
collection of peaks (bumps) and valleys (cavities).
Therefore, at the upper edge of the surface, consisting of peaks, light is mainly scattered
parallel to the direction of polarization, while at the lower edge of the surface, populated
by valleys, the main direction of light scattering is perpendicular to the polarization
direction.
Thus, the upper and lower edges of the rough surface tend to form LIPSS with mutually
perpendicular orientations.

The final shape of LIPSS results from the accumulation of absorbed energy in all surface
layers.
Therefore, the ultimate quantity of relevance for predicting light-surface coupling is the
absorbed energy $Q_\text{i}$ integrated over all $z$ layers:
$Q_\text{i}(x,y)=\int_{-\infty}^\infty Q(x,y,z) dz$.
Figure~\ref{fig:rough}(e) shows the distribution of this integrated absorbed energy
$Q_\text{i}$ where we see that the regions of high losses are mainly elongated along $x$
direction.
Therefore, we can expect that in this particular case the resulting LIPSS will be oriented
parallel to the laser polarization.
In turn, in Fig.~\ref{fig:rough}(i) we see that the spatial spectrum of $Q_\text{i}$ has a
well-recognizable shape and contains spectral features known as "type-r" and
"type-s"~\cite{Bonse2020}.
The above analysis of absorbed energy spectra across various surface layers leads to the
conclusion that type-r features primarily originate from the upper edge of the surface,
where the scattering of light is govern by surface peaks.
Conversely, type-s features predominantly arise from the lower layers of the surface, where
light scatters by the surface valleys.

\begin{figure}
  \includegraphics[width=\columnwidth]{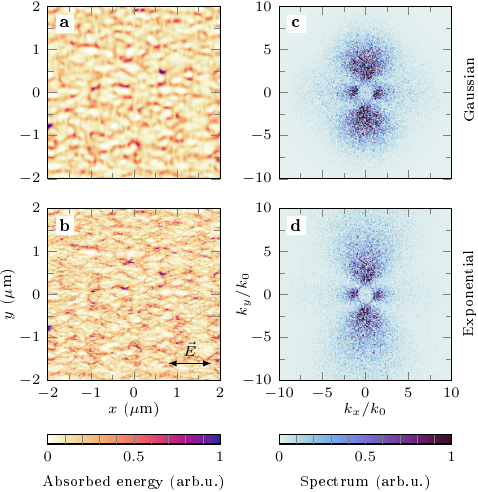}
  \caption{
    The distributions of integrated absorbed energy $Q_\text{i}$ (a,b) and their spectra
    (c,d) for rough surfaces described by the Gaussian (a,c) and exponential (b,d)
    correlation functions with the rms height $\sigma=50$~nm and correlation length
    $\xi=100$~nm.
    The double arrow shows the direction of laser polarization.
    \label{fig:corrfunc}
  }
\end{figure}

To understand how different types of correlation functions affect light absorption, let us
consider two rough surfaces with the Gaussian, Eq.~\eqref{eq:cgauss}, and exponential,
Eq.~\eqref{eq:cexp}, correlation functions.
Figure~\ref{fig:corrfunc} shows the difference between the corresponding distributions of
the integrated absorbed energy $Q_\text{i}$ and their spatial spectra.
We can see that compared to the case of the Gaussian correlation function, the distribution
of $Q_\text{i}$ at the surface with the exponential correlation function has much more fine
details.
This difference in the absorbed energy distributions is reflected in their spatial spectra,
where we see that although both spectra are qualitatively very similar, the one
corresponding to the exponential correlation function has much higher spectral components in
the $k_y$ direction.
This observation is expected since, as demonstrated earlier, the higher frequencies along
$k_y$ arise from light scattered at the peaks in the top edge of the rough surface.
As depicted in Fig.~\ref{fig:rough}, for the surface with the exponential correlation
function, these peaks are more concentrated and smaller in scale.

\begin{figure}
  \includegraphics[width=\columnwidth]{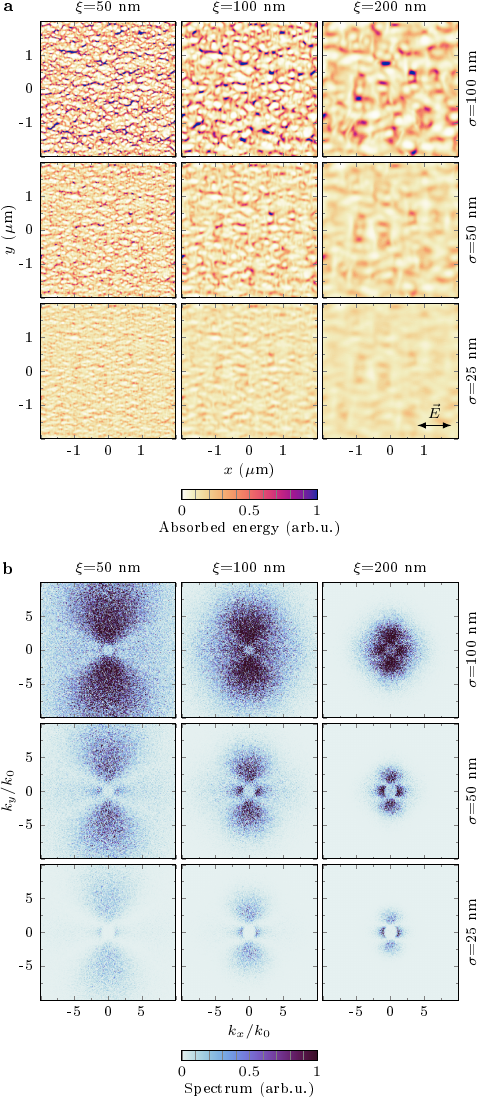}
  \caption{
    The distributions of integrated absorbed energy $Q_\text{i}$ (a) and their spectra (b)
    for rough surfaces described by the Gaussian correlation function with different values
    of the rms height $\sigma$ and the correlation length $\xi$.
    The double arrow shows the direction of laser polarization.
    \label{fig:sigma_xi}
  }
\end{figure}

Next, we consider a surface with the Gaussian correlation function and examine how changes
in its rms height $\sigma$ and correlation length $\xi$ affect the absorbed energy
$Q_\text{i}$.
Figure~\ref{fig:sigma_xi} shows the distributions of $Q_\text{i}$ and their spectra for
different values of $\sigma$ and $\xi$.
In Fig.~\ref{fig:sigma_xi}(b) we can see that the increase of the correlation length $\xi$
leads to the compression of the spectra in the $k_y$ direction.
This compression corresponds to an increase in the scale of the absorbed energy spots in the
$y$ direction (see Fig.~\ref{fig:sigma_xi}(a)).
Interestingly, at $\xi=200$~nm, the spectra become equally elongated in the $k_x$ and $k_y$
directions which means that the average sizes of the $Q_\text{i}$ spots in the $x$ and $y$
directions become the same (see the right columns in Fig.~\ref{fig:sigma_xi}(a) and (b)).
Thus, with the increase of the correlation length, we observe a transition from the
distribution of absorbed energy oriented along the laser polarization to a speckle-like
pattern.

Meanwhile, Fig.~\ref{fig:sigma_xi} shows that the increase of the rms height $\sigma$ leads
to an increase in the total absorption: the spatial distributions of $Q_\text{i}$ and their
spectra become more intense at higher $\sigma$ values.
Additionally, in Fig.~\ref{fig:sigma_xi}(b) we see that with the increase of $\sigma$ the
spectral features of type-s (see Fig.~\ref{fig:layers}(i)) gradually spread out in the $k_y$
direction.
In the case of the smallest $\xi=50$~nm and the largest $\sigma=100$~nm the type-s features
completely disappear.
With such $\sigma$ and $\xi$, the surface consists of very narrow (small correlation length)
and very deep (large rms height) inhomogeneities.
As a result, little light penetrates to the bottom of such elongated pits where it can be
scattered by the surface valleys in the right direction.

\begin{figure}
  \includegraphics[width=\columnwidth]{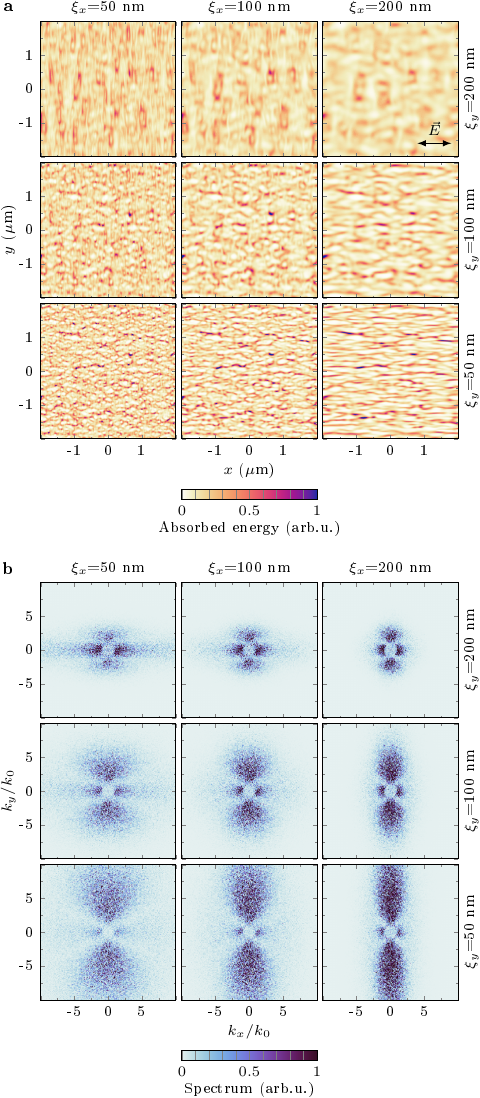}
  \caption{
    The distributions of integrated absorbed energy $Q_\text{i}$ (a) and their spatial
    spectra (b) for rough surfaces described by an asymmetric Gaussian correlation function
    with the fixed rms height $\sigma=50$~nm and different correlation lengths $\xi_x$ and
    $\xi_y$ along $x$ and $y$ directions.
    The double arrow shows the direction of laser polarization.
    \label{fig:xix_xiy}
  }
\end{figure}

Now, let us consider a surface with the Gaussian correlation function of the form
\begin{align} \label{eq:agauss}
  C(X,Y) = \sigma^2 \exp\left(-\frac{X^2}{\xi_x^2} -\frac{Y^2}{\xi_y^2}\right)
\end{align}
with the rms height $\sigma=50$~nm and two different correlation lengths $\xi_x$ and $\xi_y$
along the $x$ and $y$ directions.
Here we still use $x$-polarized incident laser pulses.
The correlation function given by Eq.~\eqref{eq:agauss} describes a surface with initially
broken symmetry where the scale of the inhomogeneities in one direction is larger then in
another.
In other words, such surface initially have a ripple-like structure in the $x$ or $y$
direction.
It may correspond to a surface subjected to mechanical polishing along one specific
direction or it may represent a surface that has already been irradiated by a small number
of preceding laser pulses.
Figure~\ref{fig:xix_xiy} shows the distributions of integrated absorbed energy $Q_\text{i}$
together with their spectra for different values of $\xi_x$ and $\xi_y$.

Let us first take a look at the case with the largest $\xi_x=200$~nm and $\xi_y$ decreasing
from 200 to 50~nm (the last columns in Fig.~\ref{fig:xix_xiy}).
This case describes a transition from a surface with isotropic roughness to one where
surface inhomogeneities, being compressed in the $y$ direction, turn out to be elongated in
the $x$ direction, parallel to the polarization.
The spectra in Fig.~\ref{fig:xix_xiy}(b) show that in this case the decrease in $\xi_y$
leads to higher spectral components in the $k_y$ direction, which in the real domain
corresponds to the compression of the absorbed energy spots along $y$.

Now, let us consider the scenario where $\xi_y=200$~nm is the largest and $\xi_x$ decreases
from 200 to 50~nm (the first rows in Fig.~\ref{fig:xix_xiy}).
Here we see a transition from the isotropic roughness to the one where the surface
inhomogeneities, being compressed in the $x$ direction, become elongated along $y$, in the
direction perpendicular to the laser polarization.
As in the previous case, the spectra in Fig.~\ref{fig:xix_xiy}(b) show that decreasing
$xi_x$ leads to the appearance of higher frequency components, but this time in the $y$
direction.
The corresponding distribution of the absorbed energy becomes elongated along $y$.

The rest of Fig.~\ref{fig:xix_xiy} confirms the previously observed trends: as surface
irregularities extend further along one of the $x$ or $y$ directions, the absorbed energy
spots become increasingly elongated in the corresponding direction.
Thus, any asymmetry in surface roughness, whether it originated as an inherent feature of
the surface or arose after laser exposure, will be magnified by subsequent laser pulses.
This mechanism constitutes the foundation of the inter-pulse feedback loop, which ultimately
leads to the formation of LIPSS from even the slightest initial symmetry breaking.
Moreover, the random nature of the surface roughness ensures that no two LIPSS patterns are
perfectly alike.
This highlights the individuality of each imprint left by exposure to light, evoking to the
distinctiveness of a fingerprint, which offers anti-counterfeiting abilities.

\begin{figure}
  \includegraphics[width=\columnwidth]{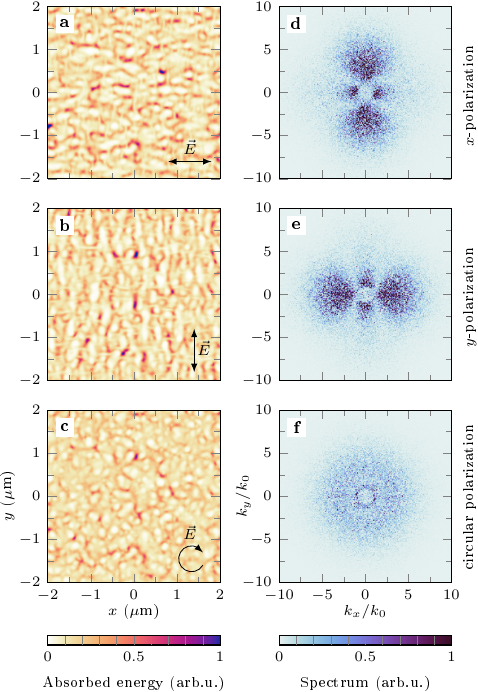}
  \caption{
    The distributions of integrated absorbed energy $Q_\text{i}$ (a,b,c) and their spectra
    (d,e,f) for $x$-polarized (a,d), $y$-polarized (b,e) and circularly polarized (c,f)
    laser pulses.
    The arrows show the direction of each laser polarization.
    \label{fig:polarization}
  }
\end{figure}

Finally, we study how the polarization of the incoming light affects the distribution of the
integrated absorbed energy $Q_\text{i}$.
Figure~\ref{fig:polarization} shows the distributions of $Q_\text{i}$ and their spectra for
$x$-polarized, $y$-polarized and circularly polarized laser pulses.
In Figs.~\ref{fig:polarization}(a,b) and (d,e) we see that in the case of $x$- and
$y$-polarized laser pulses the shapes of the $Q_\text{i}$-distributions and their spectra
are the same but rotated by 90 degrees relative to each other.
In other words, as expected, the orientation of LIPSS follows the direction of the laser
polarization.
However, Figs.~\ref{fig:polarization}(c) and (f) show that in the case of circular
polarization the distribution of $Q_\text{i}$ does not show any preferred orientation and
the corresponding spatial spectrum is symmetrical about the origin.
This can be explained by the fact that for circularly polarized light, when the electric
field vector rotates, it continuously passes through all possible orientations.
The absence of a preferred orientation in the spectral domain means that the distribution
of the absorbed energy in the real space is isotropic.
In Fig.~\ref{fig:polarization}(c) we see that the spots of the absorbed energy line up in
a circle-like pattern.
Therefore, we can expect that the resulting LIPSS will exhibit the same radial symmetry.

\begin{figure}[t]
  \includegraphics[width=\columnwidth]{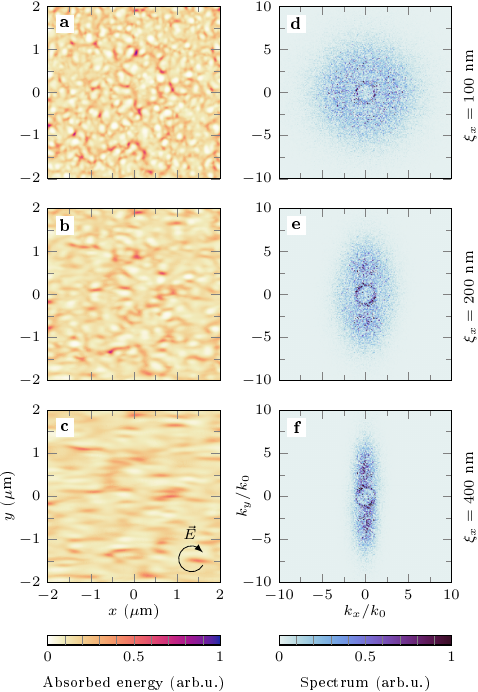}
  \caption{
    The distributions of integrated absorbed energy $Q_\text{i}$ (a,b,c) and their spectra
    (d,e,f) for a circularly polarized laser pulse interacting with rough surfaces described
    by an asymmetric Gaussian correlation function with the fixed rms height $\sigma=50$~nm
    and $y$ correlation length $\xi_y=100$~nm but with different $x$ correlation lengths:
    $\xi_x=100$~nm (a,d), $\xi_x=200$~nm (b,e) and $xi_x=400$~nm (c,f).
    The arrow show the direction of laser polarization.
    \label{fig:circular}
  }
\end{figure}

However, what happens if a circularly polarized laser pulse interacts with a surface that
initially has some asymmetry in its roughness?
In this case, the breaking of symmetry should occur not due to a dedicated direction of
laser polarization (circular polarization does not have a preferred direction), but due to
the properties of the surface itself.
In order to answer this question let us examine a rough surface with an asymmetric Gaussian
correlation function defined by Eq.~\eqref{eq:agauss}.
We fix the rms height $\sigma=50$~nm and the $y$ correlation length $\xi_y=100$~nm, and
consider the $x$ correlation length values $\xi_x=100$, 200, and 400~nm.
While $\xi_x=100$~nm defines a surface with an isotropic roughness, $\xi_x=200$ and 400~nm
values correspond to surfaces whose inhomogeneities are stretched in the $x$ direction by
two and four times, respectively.
Figure~\ref{fig:circular} shows the resulting distributions of the integrated absorbed
energy $Q_\text{i}$ and their spectra.
We can see that the increase of the $x$ correlation length leads to the elongation of the
absorbed energy spots in the $x$ direction.
Meanwhile, the corresponding spectra compress in the $k_x$ direction, remaining the same in
the $k_y$ direction.
This means that the increase of the correlation length in one direction can change the size
of the absorbed energy spots only in that same direction.
Moreover, we see that the compression of spectra in the $k_x$ direction is directly
proportional to the increase in $\xi_x$: twice larger $\xi_x$ corresponds to twice narrower
spectrum.
Therefore, by using circularly polarized laser pulses, we can effectively test the asymmetry
of surface roughness.

Interestingly, the compression of the absorbed energy spectra occurs only due to
high-frequency components, responsible for HSFLs formation.
In turn, the low-frequency components responsible for LSFLs (the central ring structures in
Figs.~\ref{fig:circular}(d--f) at frequencies close to $k_0$) do not change with the
increasing asymmetry of the surface.
Taking into account that HSFLs need ultrashort laser pulses to be formed, we can conclude
that sensing of the surface roughness asymmetries using circular polarization would be
possible only with picosecond or femtosecond laser pulses.

\section{Conclusion}
In conclusion, we use ab-initio electromagnetic simulations to study the interaction of
ultrashort laser pulses with rough surfaces.
We investigate the spatial distribution of the absorbed laser energy in order to better
understand the underlying physics of LIPSS formation.
In contrast to existing studies, to model realistic rough surfaces we apply a statistical
description of distribution of their heights.
Such approach allows us to introspect the distribution of the absorbed laser energy across
the surface selvedge.
As a result, we discovered that the high-frequency (type-s) and low-frequency (type-r)
components of LIPSS morphology emerge in distinct surface layers: type-s features
predominantly appear in the upper edge, while type-r features are mainly present in the
lower edge of the surface.
We investigate how different statistical properties of surface roughness affect the
distribution of absorbed laser energy.
Our findings demonstrate that as the correlation length increases, the pattern of absorbed
energy spots transitions from the one aligned with the laser polarization to a speckle-like
distribution.
Moreover, we demonstrate that an increase in the rms height of surface roughness leads to
the suppression of the type-r LIPSS features up to their complete disappearance.
We show that the lack of the dedicated direction of the electric field vector in circularly
polarized pulses leads to the formation of radially symmetric distribution of the absorbed
energy.
The resulting LIPSS can be expected to inherit this symmetry and form a pattern without any
preferential orientation.
The symmetry breaking induced by the roughness under circularly polarized laser pulses can
be exploited to detect and measure asymmetries in surface roughness.
In the case of circular polarization, changes in the scale of surface inhomogeneities lead
to a directly proportional deformation of the absorbed energy spectrum.
Our findings allow us to deepen the understanding of the physics behind the formation of
LIPSS and offer valuable guidance for precision laser-induced surface fabrication, with
potential applications ranging from tailored photonics devices to advanced materials design.

\begin{acknowledgements}
This work was supported by the LABEX MANUTECH-SISE (ANR-10-LABX- 0075) of Universit{\'e} de
Lyon, within the Plan France 2030 operated by the French National Research Agency (ANR).
\end{acknowledgements}


\end{document}